\documentclass[preprint2]{aastex62}
\pdfoutput=1 
\usepackage{amsmath,amstext}
\usepackage[T1]{fontenc}
\usepackage{apjfonts}
\usepackage[figure,figure*]{hypcap}

\newcommand{\angstrom}{\mbox{\normalfont\AA}}

\shorttitle{kinematics RAVE}
\shortauthors{Ramirez-Preciado V et al.}

\begin{document}

\title{Kinematic Identification of Young Nearby Moving Groups from a sample of Chromospherically Active Stars in the RAVE catalog}

\author{Valeria G. Ram\'irez-Preciado}
\affiliation{Instituto de Astronom\'ia, Universidad Nacional Aut\'onoma de M\'exico, Unidad Acad\'emica en Ensenada, Ensenada BC 22860 Mexico}
\author{Carlos G. Rom\'an-Z\'u\~niga}
\affiliation{Instituto de Astronom\'ia, Universidad Nacional Aut\'onoma de M\'exico, Unidad Acad\'emica en Ensenada, Ensenada BC 22860 Mexico}
\author{Luis Aguilar}
\affiliation{Instituto de Astronom\'ia, Universidad Nacional Aut\'onoma de M\'exico, Unidad Acad\'emica en Ensenada, Ensenada BC 22860 Mexico}
\author{Genaro Su\'arez}
\affiliation{Instituto de Astronom\'ia, Universidad Nacional Aut\'onoma de M\'exico, Unidad Acad\'emica en Ensenada, Ensenada BC 22860 Mexico}
\author{Juan Jos\'e Downes}
\affiliation{PDU Ciencias Físicas, Centro Universitario de la Región Este (CURE), Universidad de la República, 27000 Rocha, Uruguay}

\begin{abstract}

The purpose of this study is the identification of young ($1< age < 100$ Myr), nearby ($d \leqslant100$ pc) moving groups (YNMGs) through their kinematic signature. YNMGs could be the result of the recent dispersal of young embedded clusters, such that they still represent kinematically cold groups, carrying the residual motion of their parental cloud. Using the fact that a large number ($\sim$ 14000) of the RAVE sources with evidence of chromospheric activity, also present signatures of stellar youth, we selected a sample of solar type sources with the highest probability of chromospheric activity to look for common kinematics. We made use of radial velocity information from RAVE and astrometric parameters from GAIA DR2 to construct a 6-dimension position-velocity vector catalog for our full sample. We developed a method based on the grouping of stars with similar orientation of their velocity vectors, which we call the Cone Method Sampling. Using this method, we detected 646 sources with high significance in the velocity space, with respect to the average orientation of artificial distributions made from a purely Gaussian velocity ellipsoid with null vertex deviation. We compared this sample of highly significant sources with a catalog of YNMGs reported in previous studies, which yield 75 confirmed members. 
From the remaining sample, about 50\% of the sources have ages younger than 100 Myr, which indicate they are highly probable candidates to be new members of identified or even other YNMGs in the solar neighborhood.

\end{abstract}


\section{Introduction}

Young star clusters are typically found in star forming regions within Giant Molecular Clouds (GMC) while still embedded in their parental gas \citep[e.g.][]{Lada2003, Allen2007, Pelu2012}. Young stellar aggregations are usually bright in infrared wavelengths because many of their members display dusty circumstellar disks and the regions are normally associated to bright nebulosity features, all clear indicators of youth. Moreover, the dusty parental gas acts as a screen against the background population making clusters easier to spot on images. This is valid for relatively nearby associations where foreground population is minimum and contamination can be easily removed.
But once a young cluster is no longer embedded, its detection gets complicated. Typically, after 10 Myr most of the parental gas is evacuated, along with the dispersal of most circumstellar disks and the dynamical relaxation of the system, which diminishes its density \citep{Lada2003}. Then, the components of most of the clusters mix with the Galactic disk population, which challenge the identification of the cluster members.
At that evolutionary stage, emission line youth indicators such as the Li 6707\ $\angstrom$ line \citep[e.g.][]{Sicilia2005} or X-ray emission \citep[e.g.][]{Stelzel} are invoked for membership confirmation, but these diagnostics are difficult to implement in large samples. Moreover, if the cluster is close enough (10-$10^2$ pc), it cannot be distinguished as an overdensity in the sky because it covers very large areas.

A feature that can help to identify a dispersed cluster is the common kinematics still imprinted in their members; at ages of one to a few tens of Myr, young groups still stand out from the local velocity ellipsoid and have small velocity dispersions \citep{Antoja2012}. Known at this point as Young Nearby Moving Groups \citep[YNMG; for a detailed review on the topic see][]{Torres2008}, they can still be identified as they keep moving together away from their parental cloud before entering the Galactic Disk highway.\\
 
The motivation of this work is to present a new method and procedure to identify emerging groups of young stars in the disk of the Galaxy through their kinematic signature. Finding such groups can be useful to understand the early evolution of unbound stellar groups, particularly how they disperse and integrate to the population of the galactic disk. Also, we look for a suitable technique for the identification of YNMG members, taking advantage of the increasing availability of data to provide the necessary parameters to construct full position-velocity vectors. Moreover, studying these kind of stars in the Solar Neighborhood can contribute to the understanding of star formation on the disk, as well as to its kinematic evolution.
 
Current identification methods are largely based on proper motions \citep[e.g.][]{Aguilar1999, Gagne2018} and positions \citep[e.g.][]{kop2018}, but this limited kinematical information makes group identification inconclusive, or unreliable.
\cite{Riedel2017} developed a statistical method for the identification of YNMGs through their position and partial kinematical information, but as we mentioned before, position is not a reliable parameter for the identification of such disperse groups of stars.
That is why it is necessary to implement other methods for its identification.

In this paper we present a simple and effective method to identify candidate members of YNMGs when their full 6D position-velocity vectors are known. The method can be applied to any sample that combines reliable proper motions, radial velocities and distances. Our test dataset is the RAdial Velocity Experiment (RAVE) catalog \citep{kunder2017}, known to contain solar-type young star candidates. 

This paper is organized as follows: we describe our sample selection in section 2, followed by a description of our methology in section 3. The results of applying our method to the selected dataset are presented in section 4. Finally, section 5 contains a discussion and summary of this work.

\section{Sample \label{s:sample}}

The RAVE catalog contains stars with F, G and K spectral types, distributed in brightness within 8 and 12 magnitudes in the near-infrared $I$ band across the Southern hemisphere sky.
The current (Data Release 5) RAVE catalog contains parameters for 520,630 individual stellar spectra, which are classified by means of a method called Local Linear Embedding \citep[LLE,][]{Matijevic2012} which is based on a dimensionality reduction algorithm \citep{LLE}. 
The classification consists on applying the method directly to stellar spectra, reducing the number of dimensions (defined from a set of spectral features across previously defined spectral bins) that are needed to make the classification of a certain type of star. These dimensions are defined as spectral features common to an specific type of source.
Each observed spectrum is compared against a grid of synthetic spectra in order to define a comparison sub-sample (5000 stars) from the previous data release. If the observed spectrum presents, with a certain level of confidence, features similar to those of other previously classified spectra in one or more dimensions, it is assigned a ``flag''  value \citep{Matijevic2010} in the corresponding dimension, indicating the nearest classification type. For each source listed, twenty flags representing the spectral feature dimensions are listed and each flag contains a letter representing the closest among eleven different stellar spectral classes of sources \citep{Matijevic2012}. The first three flags are those that have the highest weight in the classification. If these three first flags coincide, then the star has a high probability of belonging to that class. 

For our study, we were interested in chromospherically active stars (CAS) in the RAVE catalog because it is estimated that about 40\% of the CAS in RAVE coincide with high H$\alpha$ emitters from the ESO-GAIA catalog \citep{Zerjal2013,zwitter2015}. The other $\sim$60\% left could be young stars with not strong emission in that line or could be contaminants such as giants (see Section \ref{s:diagrams-HR}).
For the selection of our sample we considered stars for which the 3 first flags coincided with type \textit{``chromospherically active''}. With this criterion we obtained a sample of 3128 stars with a high level of confidence of being CAS, from which many of them are expected to be young stars. 

\subsection{GAIA DR2 kinematic parameters}

The RAVE catalog provides radial velocities for all sources in our sample {RAVE provides good measurements for radial velocities, with uncertainties of 5 km/s or less}). In order to obtain the kinematic parameters for our main analysis, we complemented the RAVE CAS sample with astrometric parameters obtained from the GAIA Collaboration Data release 2 \citep[hereafter GAIA DR2][]{GAIADR2}. From GAIA DR2 we obtained RA and DEC positions, parallaxes and proper motions in RA and DEC, for all the sources in the sample. Using the \texttt{TOPCAT} tool version 4.6-1 \citep{topcat}, we applied the method of \citet{bj2018} to convert parallaxes into distance estimates using a local density exponential decrease prior with a scale parameter $h=500$ pc. With this information we were able to provide 6-dimensional position-velocity vectors for all the sources in the sample.

\section{Data Analysis\label{s:analysis}}
\subsection{Velocity ellipsoids projections \label{s:analysis:ss:kinematics}}

The velocity vector of a star is fully determined by the radial velocity and the two components of its tangential velocity on the celestial sphere in addition to a parallax or distance to the source. In the 6-dimensional position-velocity space (hereafter PV6), each star is defined by six observational parameters: two coordinates (e.g. $l,b$ in the Galactic coordinate system), the corresponding two proper motion components $\mu_{l}$, $\mu_{b}$, a parallax $\varpi$ from which a distance can be estimated, and the radial velocity, usually calculated from Doppler shift measurements on a spectrum. 
From this, a PV6 vector ($X$, $Y$, $Z$, $U$, $V$, $W$) can be determined.

\begin{figure}[htbp]
\centering
\includegraphics[width=0.5\textwidth]{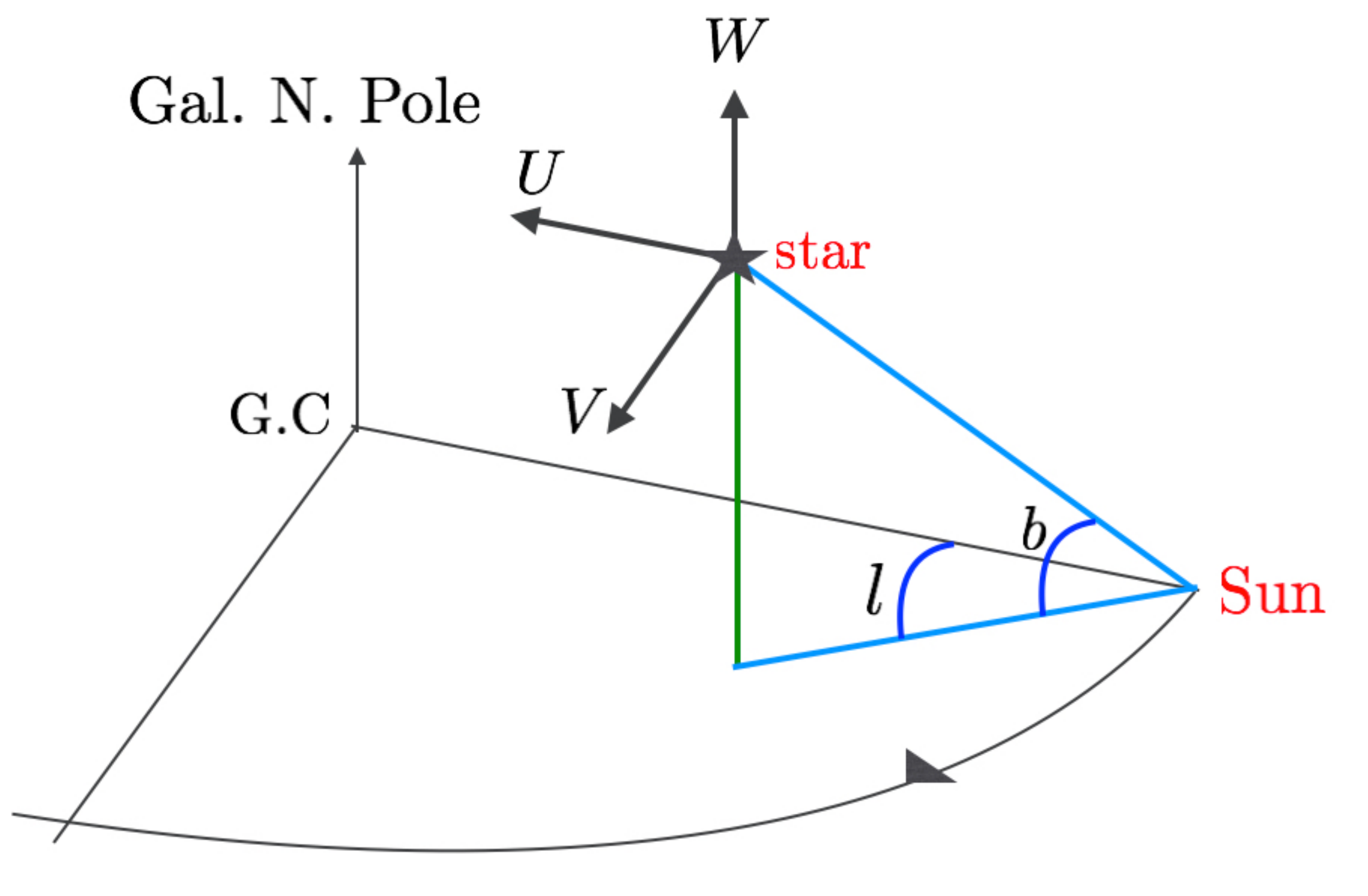}
\caption{Graphical description of the $(U,V,W)$ coordinates frame according to the galactic coordinates as expressed in equations \ref{u1}, \ref{u2} y \ref{u3}.}
\label{schon12}
\end{figure}

We use a heliocentric frame and following the canonical scheme \citep[e.g.][]{schonrich2012}, $X$ and $U$ are positive toward the galactic center, $Y$ and $V$ are positive along the direction of the galactic rotation, while $Z$ and $W$ are positive toward the galactic north pole. $U$, $V$ and $W$ are defined by equations \ref{u1},\ref{u2} and \ref{u3} below (see Figure \ref{schon12}, based on Figure 1 of \citeauthor{schonrich2012} 2012): 

\begin{align}
\label{u1}
U = v_r (\cos{l} \cos{b}) - v_l (\sin{l}) - v_b (\sin{b} \cos{l})\\
\label{u2}
V = v_r (\sin{l} \cos{b}) + v_l (\cos{l}) - v_b (\sin{b} \sin{l}) \\
\label{u3}
W = v_r (\sin{b}) + v_b (\cos{b})
\end{align}
where $v_r$ is the radial velocity, $v_l$ and $v_b$ are the tangential components of the velocity.\\

The U, V, W velocity components give us the velocity distribution for a given population, usually described to first approximation as a \textit{velocity ellipsoid}, and scatter plots of any two such components give us projections of this ellipsoid.
These projections can be used to look for overdensities that may represent YNMG of stars \citep[e.g.][]{Antoja2009,Antoja2012}. 

\subsection{Lack of structure in the velocity ellipsoid projections}

\begin{figure*}[htbp]
	\centering
	\includegraphics[width=0.46\linewidth]{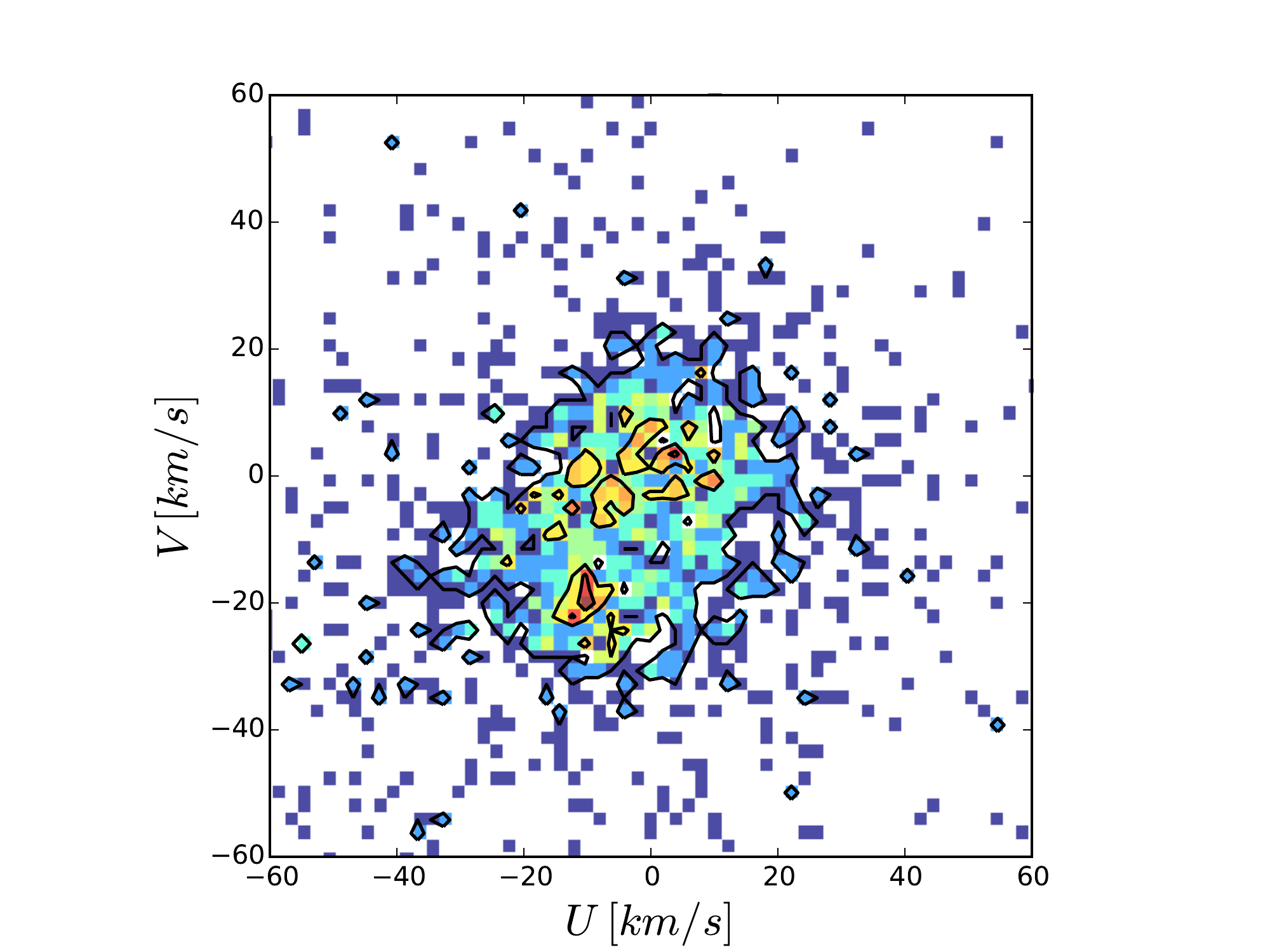}
	\includegraphics[width=0.46\linewidth]{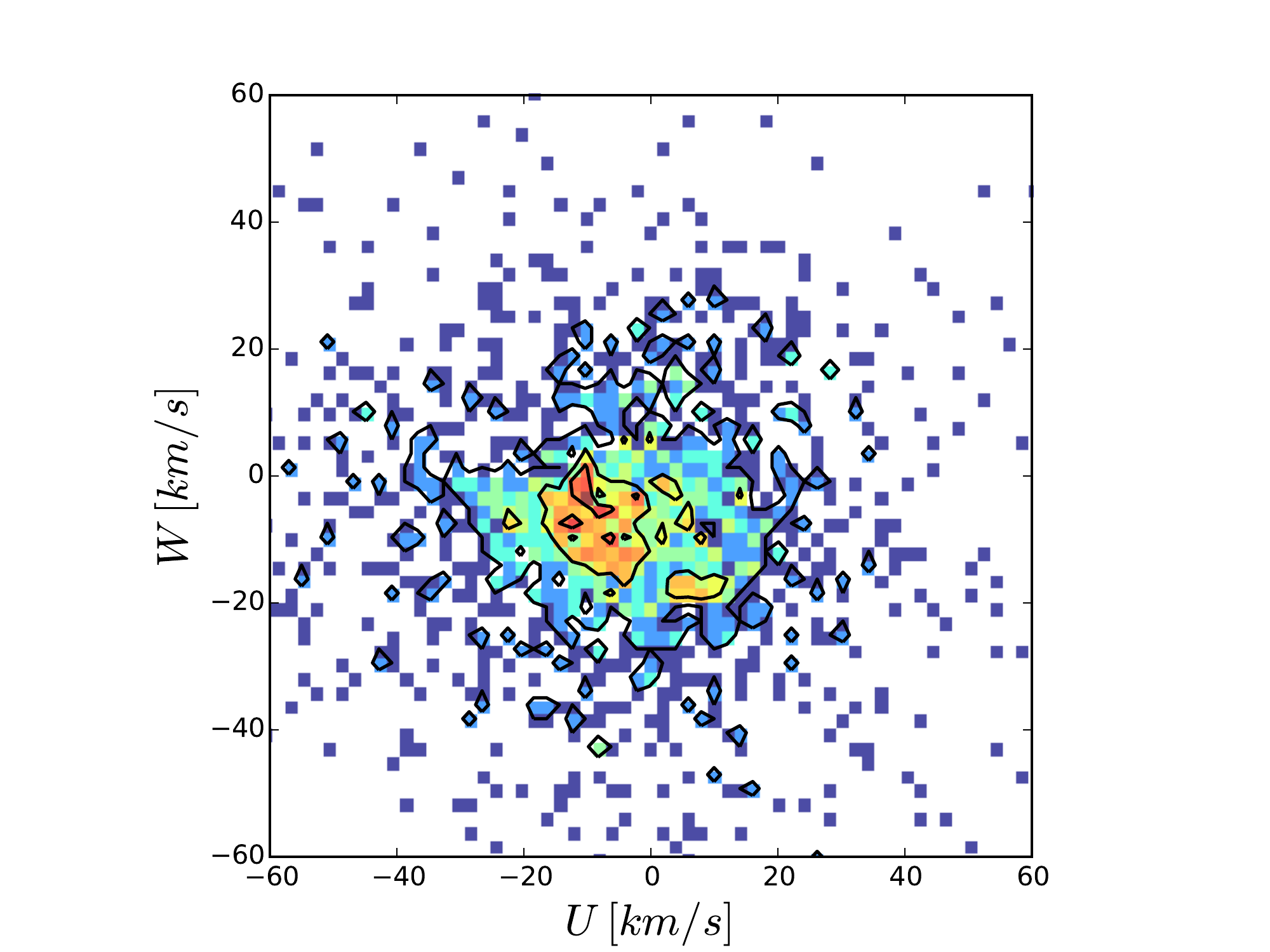}
	\includegraphics[width=0.46\linewidth]{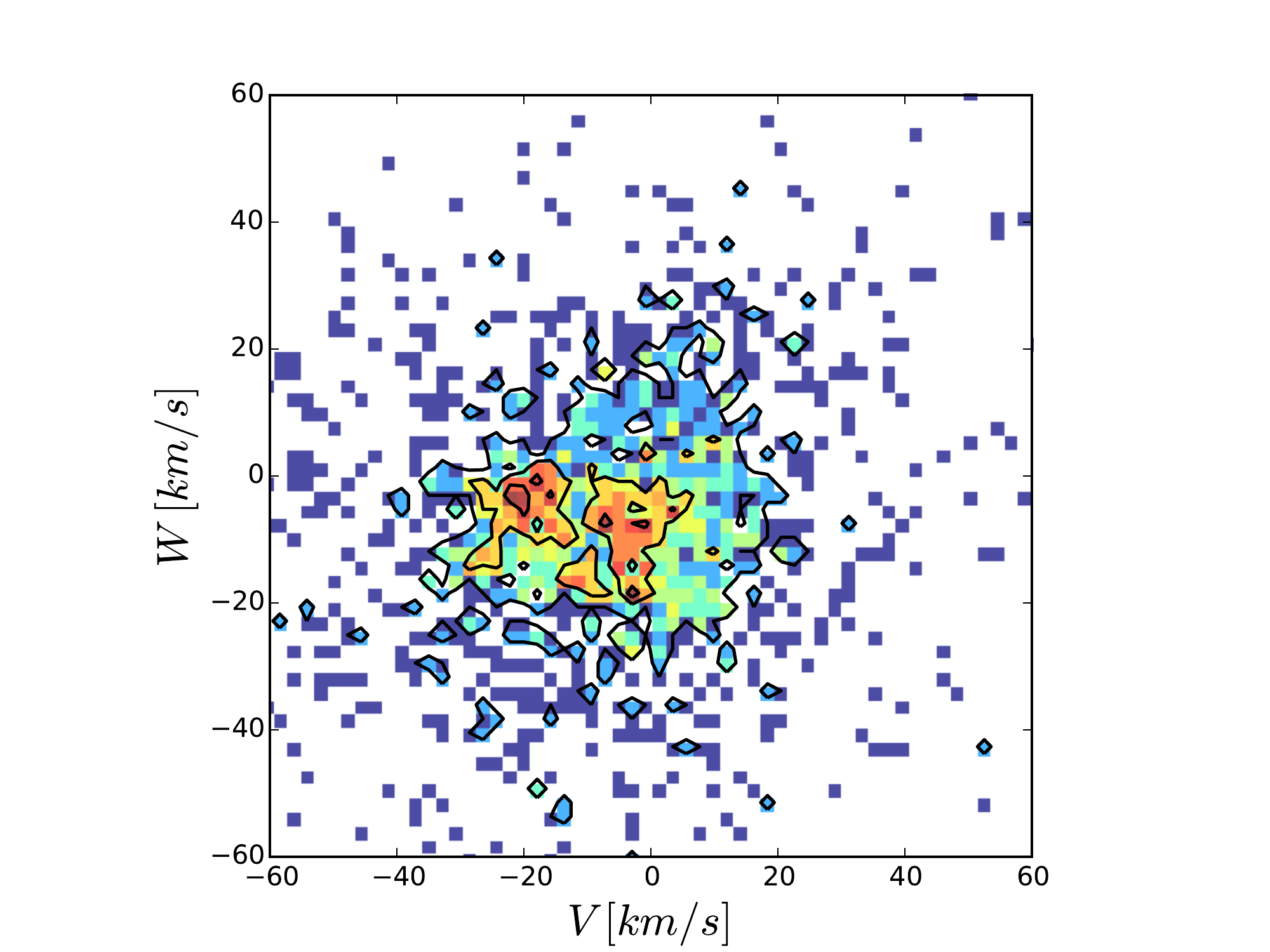}
	\includegraphics[width=0.072\linewidth]{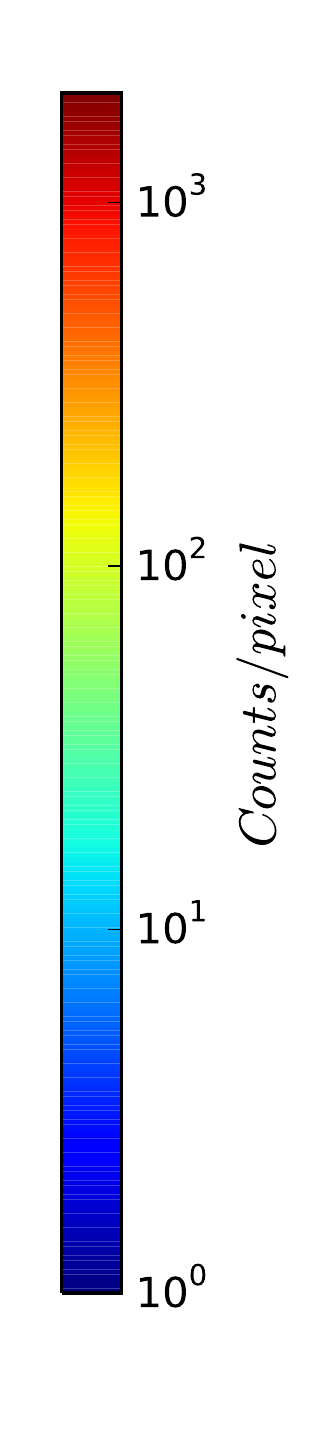}
	\caption{The panels show the UV (\textit{left}), UW (\textit{right}) and VW (\textit{bottom}) projections of the velocity ellipsoid for the RAVE sample of CAS. The resolution of the 2D histogram is 3 km/s. The color bar indicates density of stars per pixel in a logarithmic scale. The isocontours trace the most prominent overdensities in the kinematic space.}
	\label{UV-maps}
\end{figure*}

We constructed the velocity ellipsoid projections (see Figure \ref{UV-maps}) for our sample of RAVE CAS using 2-dimensional histograms. We look for substructure on top of the smooth ellipsoidal distribution, as signal of moving groups.
Our criteria for these plots were \textit{a)} a resolution of 3 km/s per bin, and \textit{b)} to consider only those stars with velocity moduli smaller than 600 $\mathrm{km/s}$. The latter was chosen from a histogram for $(U,V,W)$ values, indicating that more than 80\% of our original sample was smaller than 200 $\mathrm{km/s}$.

All three projections show an ellipsoidal-like distribution for our sample. The left panel of Figure \ref{UV-maps} shows a clear vertex deviation with an angle of approximately 15 $^\circ$, and an indication of an overdensity near $(U,V)=(-15,-20)$ plus a few small lumps near the center. The $VW$ and $UW$ projections are mostly featureless. However, we conclude that the signal to noise of these histogram images may not be high enough to resolve substructure clearly with our relative small sample. \\
Increasing the resolution of these 2D histograms did not improve our results either, because the noise level also increased, making it actually more difficult to distinguish any overdensity features.

\begin{figure}[htbp]
\centering
\includegraphics[width=0.54\textwidth]{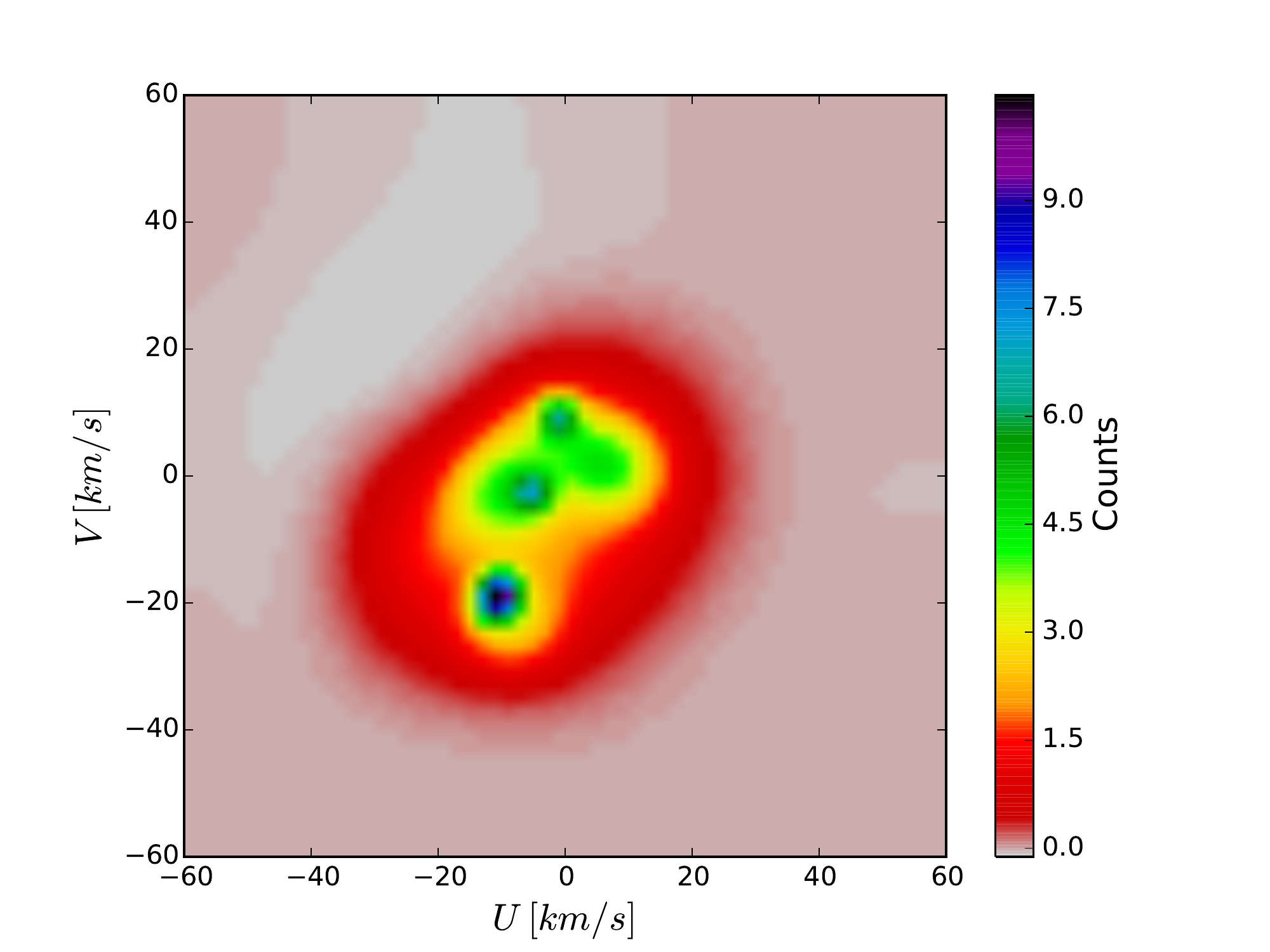}
\caption{Density map of the UV projection of the velocity ellipsoid for RAVE CAS after applying the wavelet filtering process based on MVM. Although the substructure of left panel in Figure 2 is more evident here, the resolution of this image (3 km/s) is still not sufficient to carry out a satisfactory separation of individual moving groups.The colorbar indicates source density.}
\label{wave}
\end{figure}

Figure \ref{wave} shows a further effort to highlight and identify structures within the $UV$ projection. We maintaned the resolution to a projection bin of 3 km/s and we applied a wavelet filtering method to the resultant image in order to highlight overdensity regions while removing extended low density structure.
For this purpose we used an algorithm (B. Vandame, personal communication) based on the Multi-scale Vision Model (MVM) by \citet{Rue1997}. 
This filtering process allowed to highlight some possibly significant over-densities in the UV projection, as shown in Figure \ref{wave}.
Our wavelet image shows four main lumps with sizes of about 10 km/s.
 The sizes and the separations between groups in our wavelet filtered map are consistent with the velocity ellipsoid projections for moving groups in the solar vicinity by \citet{Antoja2012}.  The largest lump near (U,V)=(-10,-20) is actually close with the Hyades, and the central lump is close to Coma Berenice,  as reported in that work, but the coincidences are not exact, and the other two small lumps in our diagram are not directly related to any of their groups. On one hand, the differences could be explained by our use of GAIA DR2 parameters, which may be refining some values and  highlighting distinct features. But on the other hand, our CAS sample is very different from theirs and we cannot venture to claim a real coincidence. Moreover, on Antoja work, they also used a higher resolution in their maps, but we cannot reach that resolution due to the size of our sample, which complicates the use of the velocity projections to identify additional structure. 
 As we expect more than four YNMGs in our sample, we implemented an additional method to try to identify them.

\subsection{The Cone Method \label{s:analysis:ss:cone}}

As shown in the velocity ellipsoid projection figures (Figures \ref{UV-maps} and \ref{wave}), some substructure becomes apparent when using some contrast enhancing techniques, like the wavelet filter. The basic problem we have here is that we are dealing with projections, which diminishes the contrast of 3D substructure. As we are searching for stars that belong to a YNMG,they should constitute
a kinematically ``cold'' group. This means that their velocity vectors, when seen from a reference frame away from its own barycenter, should all point roughly in the same direction. This is the basis of the method we introduce here. While this is still based on a projection, it is entirely different in its construction.

\begin{figure}
	\centering
	\includegraphics[width=0.9\linewidth]{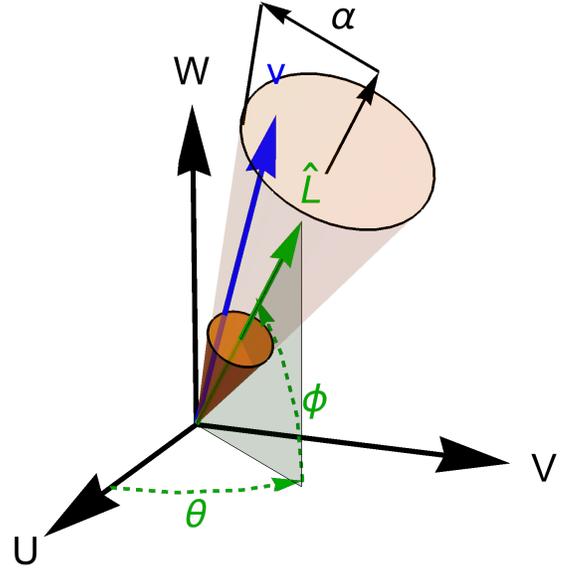}
	\caption{The panel shows the configuration of the cone on each star. It shows how the direction of the velocity vector fall within the cone. Thus, all stars that have the same configuration of $\theta$ and $\phi$ are grouped.} 
	\label{cone-method}
\end{figure}

First, a cone with vertex at the velocity frame origin is defined (see Figure \ref{cone-method}). The angles $\theta$ and $\phi$ define its orientation in this space ($0 < \theta \leqslant 360^\circ$ along the $U$-$V$ plane, measured from the first toward the second axis; $-90^\circ \leqslant \phi \leqslant 90^\circ$ perpendicular to the first axis and null for $W = 0$) and the angle $\alpha$ denotes its aperture. The unit vector $\hat{L}$ indicates the cone symmetry axis:

\begin{equation}
\label{cono2}
\hat{L}=[\cos{(\theta)} \cos{(\phi)}, \sin{(\theta)} \cos{(\phi)},\sin{(\phi)}].
\end{equation} 

Then, we identify all stars in the sample whose velocity vector $\vec{v}$ falls within this cone, i.e. the following condition is satisfied:
\begin{equation}
\label{constriction}
\frac{\hat{L} \cdot \vec{v}}{|\vec{v}|} < \cos{(\alpha)}.
\end{equation}

Notice that this condition is set in velocity space, not in configuration space, which means that the star position in configuration space is irrelevant.\\
All stars that fall within the cone are assigned to this particular combination of $(\theta, \phi)$ values. We then establish a $(\theta, \phi)$ grid on the unit sphere and compute the number of stars assigned to each such gridpoint, calling it the Cone Method Sampling (CMS). The grid is designed so that each cell subtends equal solid angles. We can then generate star count maps on the unit sphere
where local peaks indicate groupings of stars that share the same velocity vector orientations (within an angle $\alpha$). Reducing the cone opening makes the criterion more stringent,
allowing to identify colder clumps, but reduces the number of stars within a group. It is also obvious that $\alpha$ should not be reduced below the uncertainty in velocity orientations
resulting from the errors in the observables.\\
After we identified substructure using this procedure, it is necessary to assign to each peak a statistical significance, i.e. a measure of its probability that it is not a mere chance fluctuation. For this, we need a null-hypothesis (NH) that all substructure is merely due to fluctuations due to counting of discrete events in a mesh. In our case, we use as NH a trivariate Gaussian distribution whose centroid and extent are given by the individual means and standard deviation of the individual $U$, $V$ and $W$ distributions of the sample. No correlation between the individual components is assumed, so no vertex deviation is considered. We then performed Monte Carlo sampling of the NH to construct 5,000 synthetic samples of equal size to the real one. At the end we built $(\theta, \phi)$ maps of the expected median density of sources at each gridpoint under the hypothesis that no substructure really exists. We used these values, gridpoint by gridpoint, to establish the statistical significance of the peaks in the real sample, as we describe below.

\begin{figure*}
\centering
\includegraphics[width=0.65\textwidth]{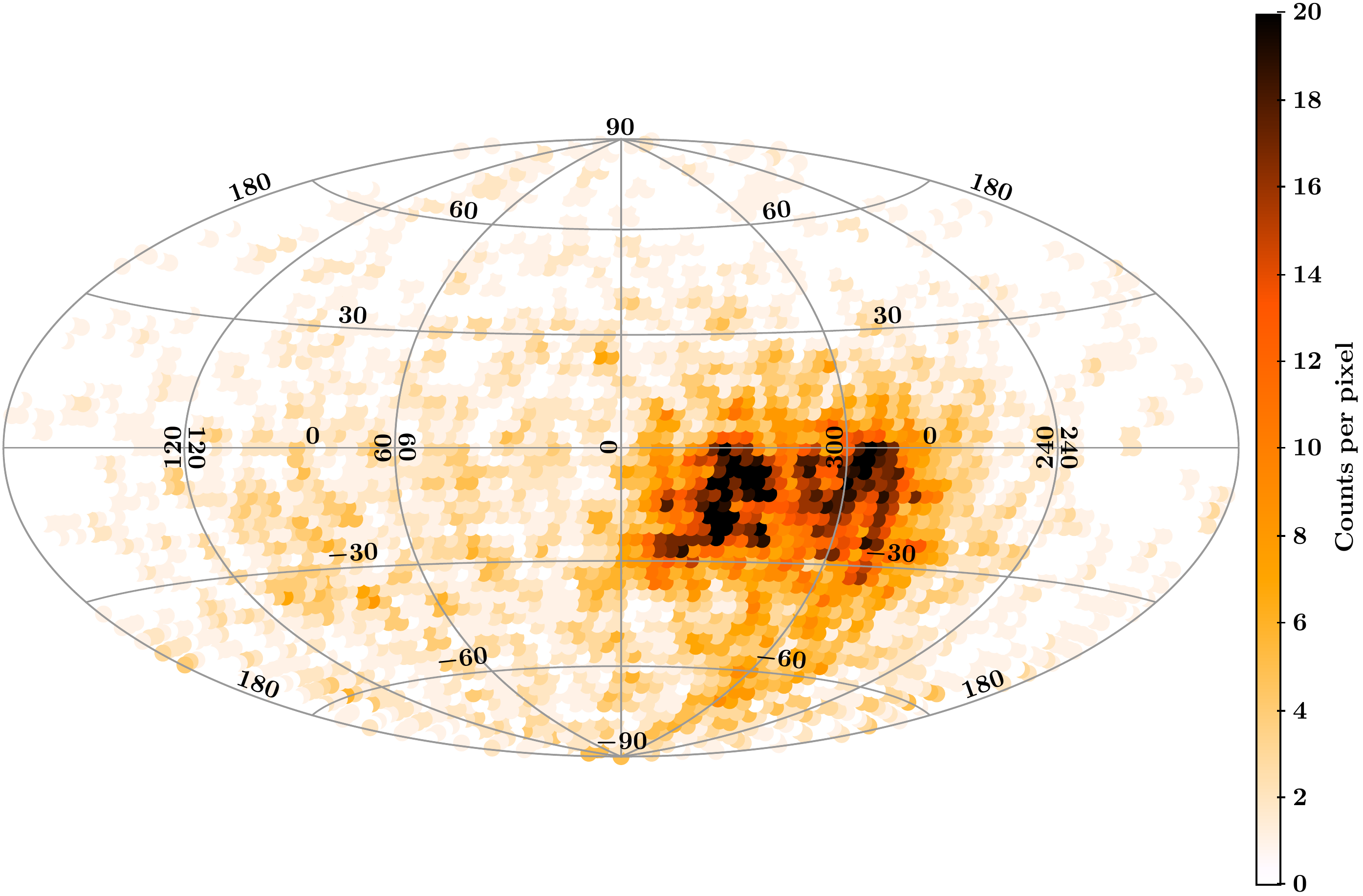}
\caption{The panel shows the counts per cone in the $(\theta, \phi)$ grid after applying the CMS to the RAVE CAS sample. This is a molleweide projection and shows areas of accumulation of sources with common kinematics. The colorbar indicates density levels in this space as counts per pixel.}
\label{ncount}
\end{figure*}

\subsection{Structure separation}
We applied the CMS, as previously described (see Section \ref{s:analysis:ss:cone}), to our sample of RAVE CAS, in order to obtain the distribution of stars that satisfied the dot product condition in inequality \ref{constriction}. 
For our purposes, we chose the value $\alpha=3^\circ$, and our grid was constructed as follows: first, we made a $\cos(b)$ correction on latitude, to assure uniform coverage; then we used Nyquist sampling in order to reduce the step to half the resolution of the grid so no stars are left out of the counting.
A map of the distribution of our CAS sample in the $(\theta, \phi)$ grid is shown in Figure \ref{ncount}.

\begin{figure*}[ht]
	\centering
	\includegraphics[width=0.45\linewidth]{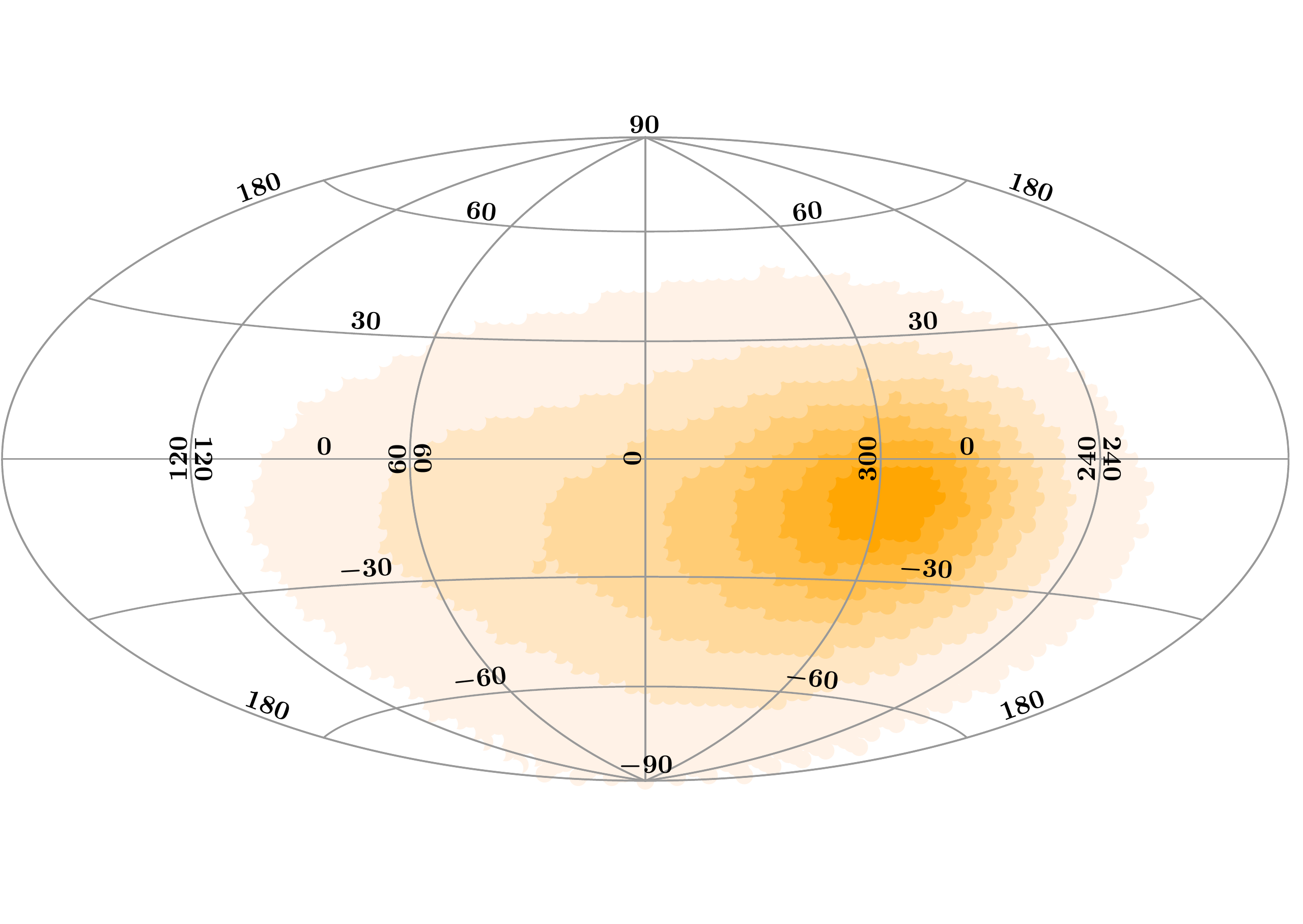}
	\includegraphics[width=0.5\linewidth]{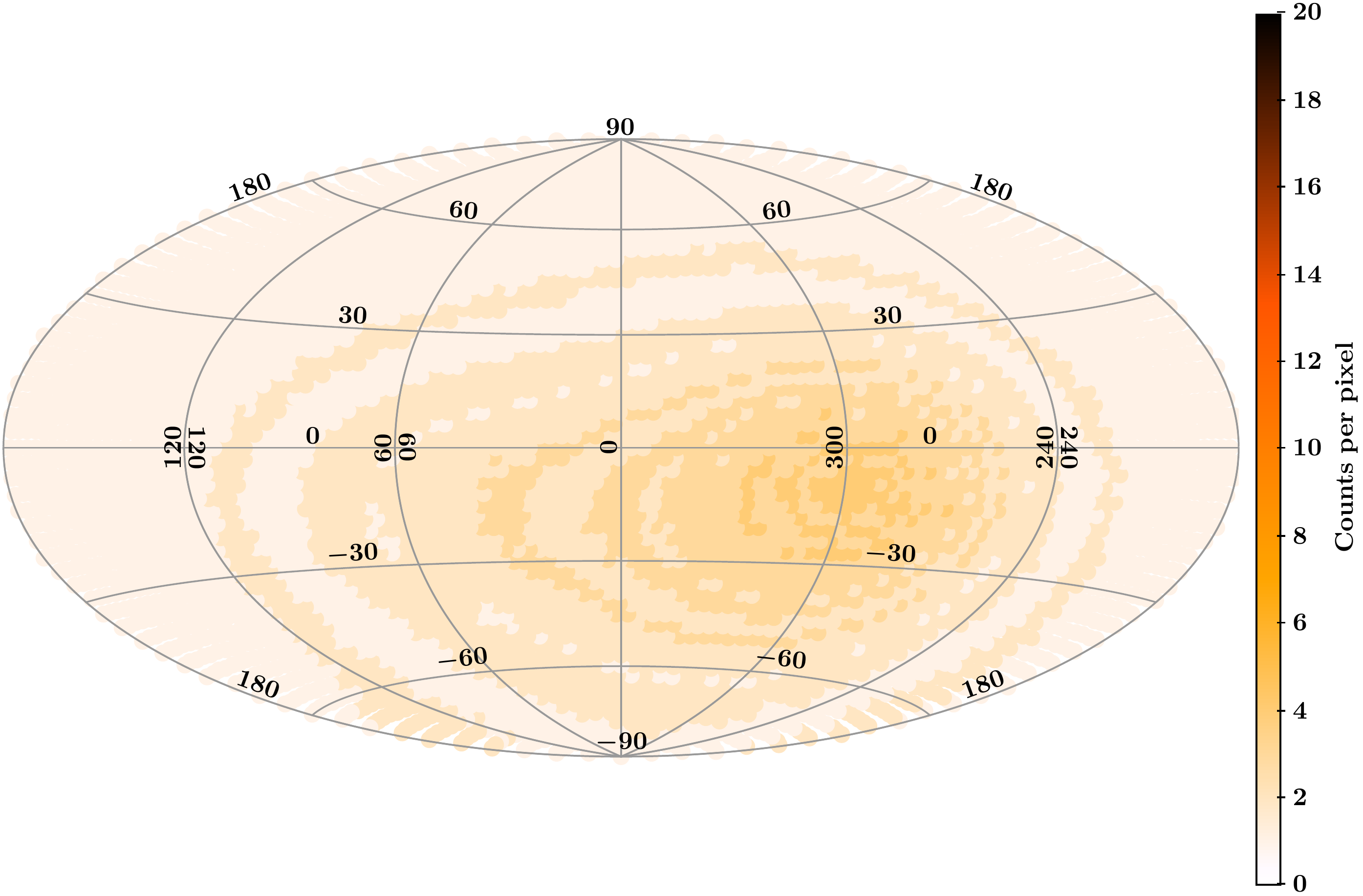}
	\includegraphics[width=0.5\linewidth]{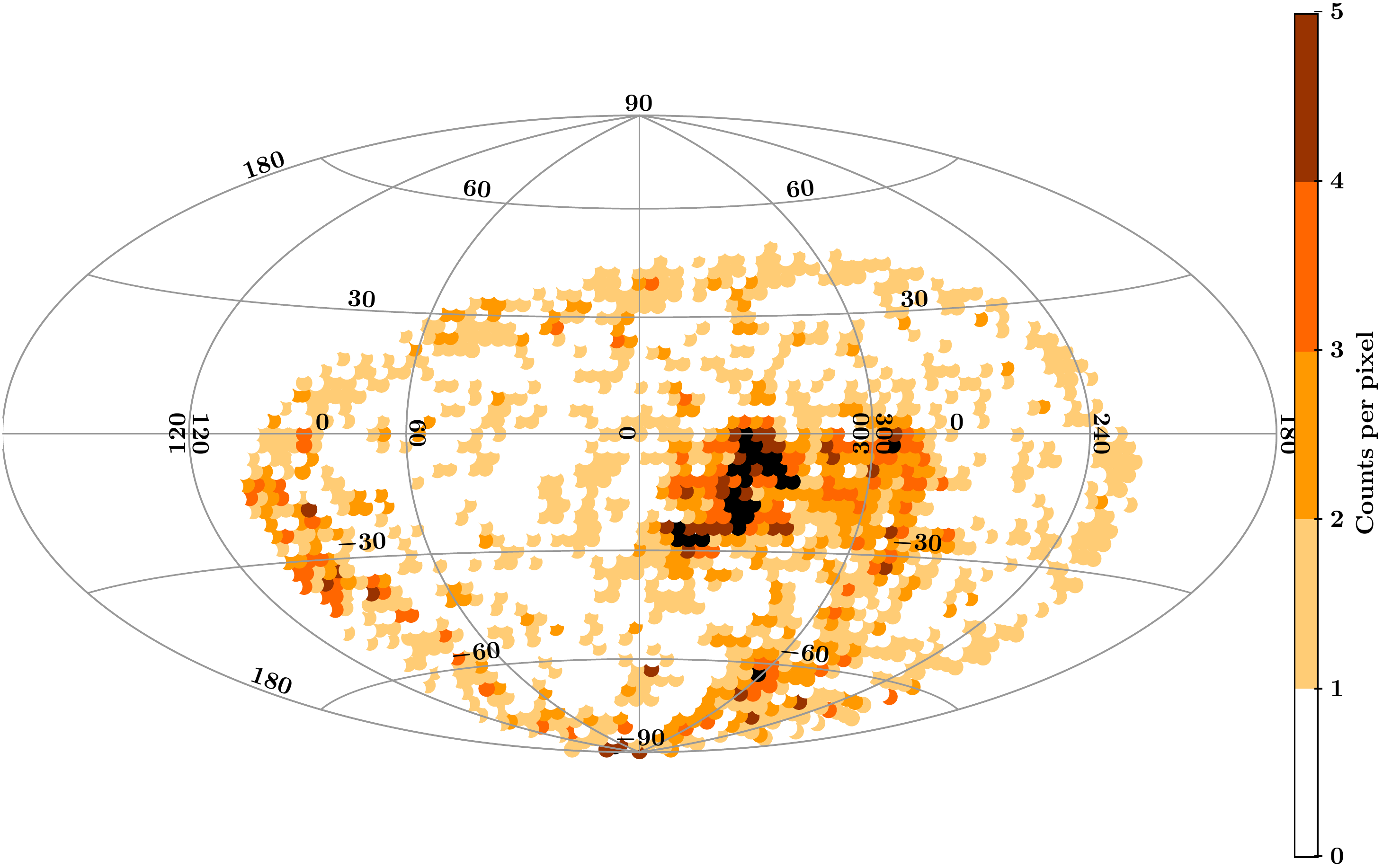}
	\caption{The panel shows the statistics calculated from the sample of stars. The (\textit{left}) figure shows the median of the sample. The (\textit{right}) figure shows the deviation and the \textit{bottom} shows the statistical significance. The three color bar denote density per pixel of stars.}
	\label{med-dev}
\end{figure*}

In order to obtain a more reliable identification of possible YNMG from the CMS, we developed and implemented a method to identify structures within the map described above, based on the statistical significance of the counts at each position.\\
Using the Monte Carlo realizations, we used the median of the count density at each position in the mesh for the NH case as a central estimator.
We used the median because it is a more robust indicator for the distributions in each mesh point, which are in most cases positively skewed. Then, we obtained a deviation by calculating the absolute difference between the median and the value corresponding to the 90 percentile, also at each position of the mesh. The corresponding median and deviation maps are shown in the top-right and top-left panels of Figure \ref{med-dev}.

We define a \textbf{reliability} range at each position in the mesh as:
\begin{equation}
S = \frac{observed \ data \ - \ median \ (NH)}{deviation \ (NH)}
\label{signif}
\end{equation}

This way, we define high significance on all points as $S > 1.5$ and low significance as $1<S<1.5$.\\
A map that shows the final distribution of CMS counts using the significance estimator, is shown in the bottom panel of Figure \ref{med-dev}. We found a total of 646 stars located in mesh positions with high significance. These sources represent a new sample of solar type CAS candidates to be members of recently disaggregated young star clusters.  

\subsection{Uncertainty cone}

It was necessary to corroborate that our choice for the cone opening value, $\alpha$, was adequate for the detection of groups with similar velocity vector orientation in the RAVE CAS sample. This is, we need to make sure that $\alpha$ is consistently wide enough compared to the uncertainty in the angle between $\hat{L}$ and $\vec{v}$. For this purpose, we constructed an error cone from the uncertainties in the $U$, $V$ and $W$ components.
Using the following transformation expressions from the velocity to the cone space:

\begin{equation}
\label{conoerror11}
\theta = \arctan{(\frac{W}{\sqrt{U^2 + V^2 }})},
\end{equation}

\begin{equation}
\label{conoerror21}
\phi = \arctan{(\frac{V}{U})},
\end{equation}

we determined the error propagation to first order, which provides $\delta \theta$ and $\delta \phi$. This way we can determine $\delta \alpha$ as the equivalent radius of the area $\delta \theta \times \delta \phi$, which corresponds to the opening of the cone formed with the uncertainties. A detailed derivation of $\delta \alpha$ is included on Appendix \ref{App1}.
If $\delta\alpha$ is consistently smaller than the characteristic aperture $\alpha$ of the map (in our case $3^\circ$.), we can trust that the aperture used is adequate for finding common kinematics between stars in our sample. We confirmed that more than 90\% of the stars in our sample have an opening of the uncertainly cone smaller than $\alpha$, indicating that the $\alpha$ value we use is reliable.

\section{Results \label{s:results}}

\subsection{Moving Groups \label{s:mg}}


After analyzing the significant regions with the method described above, we need to know if our selected stars have been already identified as  members of YNMGs. For this purpose, we cross matched our list with a compilation of known members of YNMG including results from: \citet{Torres2006, Torres2008} ($\beta$ Pictoris, Columba, Tucana-Horologium), \citet{Dawson2012, Dawson2013} ($\eta$ Chameleontis), \citet{Ducourant2014} (TW Hydrae), \citet{Elliott2014} (AB Doradus, Argus, $\beta$ Pictoris, Carina, Columba, $\eta$ Chameleontis, Octans, TW Hydrae), \citet{Galvez2010, Galvez2014} (Castor), \citet{Malo2013, Malo2014} ($\beta$ Pictoris, TW Hydrae, Tucana-Horologium, Columba, Carina, Argus, AB Doradus), \citet{Riedel2014} ($\eta$ Chameleontis, TW Hydrae, $\beta$ Pictoris, Octans, Tucana Horologium, Columba, Carina, Argus, AB Doradus), \citet{Gagne2015, Gagne2018} (Argus, Columba, $\beta$ Pictoris, AB Doradus, Carina, TW Hydrae, Tucana Horologium), \citet{Desilva2009, Desilva2013} ($\eta$ Chameleontis, TW Hydrae, $\beta$ Pictoris, Octans, Tucana-Horologium, Columba, Carina, Argus, AB Doradus), \citet{Cruz2009} (AB Doradus, $\beta$ Pictoris, TW Hydrae), \citet{Kraus2014} (Tucana-Horologium), \citet{Makarov2000} (Carina), \citet{Moor2013} (Columba, Carina, Argus, AB Doradus, $\beta$ Pictoris), \citet{Murphy2015} (Octans), \citet{Shkolnik2012} (AB Doradus, $\beta$ Pictoris, Carina, Castor, $\eta$ Chameleontis, Columba, TW Hydrae, Tucana-Horologium) and \citet{zuckerman2004} (AB Doradus, $\eta$ Chameleontis). The catalog contains information for over 2300 members of associations at distances smaller than $\sim200$ $\mathrm{pc}$, with revised positions and information about which YNMG they each belong. We found a total of 75 matches with 10 known YNMGs, which are listed in Table 1 of Appendix \ref{App3}. In Figure \ref{moving-groups} we show a map, in Galactic coordinates, of these stars. The remaining 571 sources have no matches with recent literature on YNMG.

To corroborate the reliability of our technique, we made the same analysis for a sample of 275 stars from BANYAN IV \citep{Malo2014}. We found that with our method we recovered 173 stars of the initial sample which have $S>1.5$ and 270 sources with $S>1$. 
Determination of membership for each candidate using typical methods \citep[e.g. spectroscopy; see for instance][]{Binks2015} is beyond the scope of this study. 
In the ideal case, we should produce a table with individual membership probabilities based on known properties of YMNG groups, like average radial velocity, average distance, etc. However, this kind of information is difficult to recollect in a consistent way for most YNMGs. We only were able to make an acceptable comparison with the Beta Pictoris group \citep[distance=18-40 pc, $\mathrm{<v_r>=60}$km/s, extension = 40 pc.][]{Malo2013, Moor2013}, were 80 stars in our remaining candidate sample coincide within the uncertainties with the characteristics of this group. 
For this reason, we chose instead to analyze the location of our CMS significant CAS in the HR diagram, focusing on the ages of the candidates, in order to highlight possibly young sources. \\

\begin{figure*}[h]
	\centering
	\includegraphics[width=0.8\linewidth]{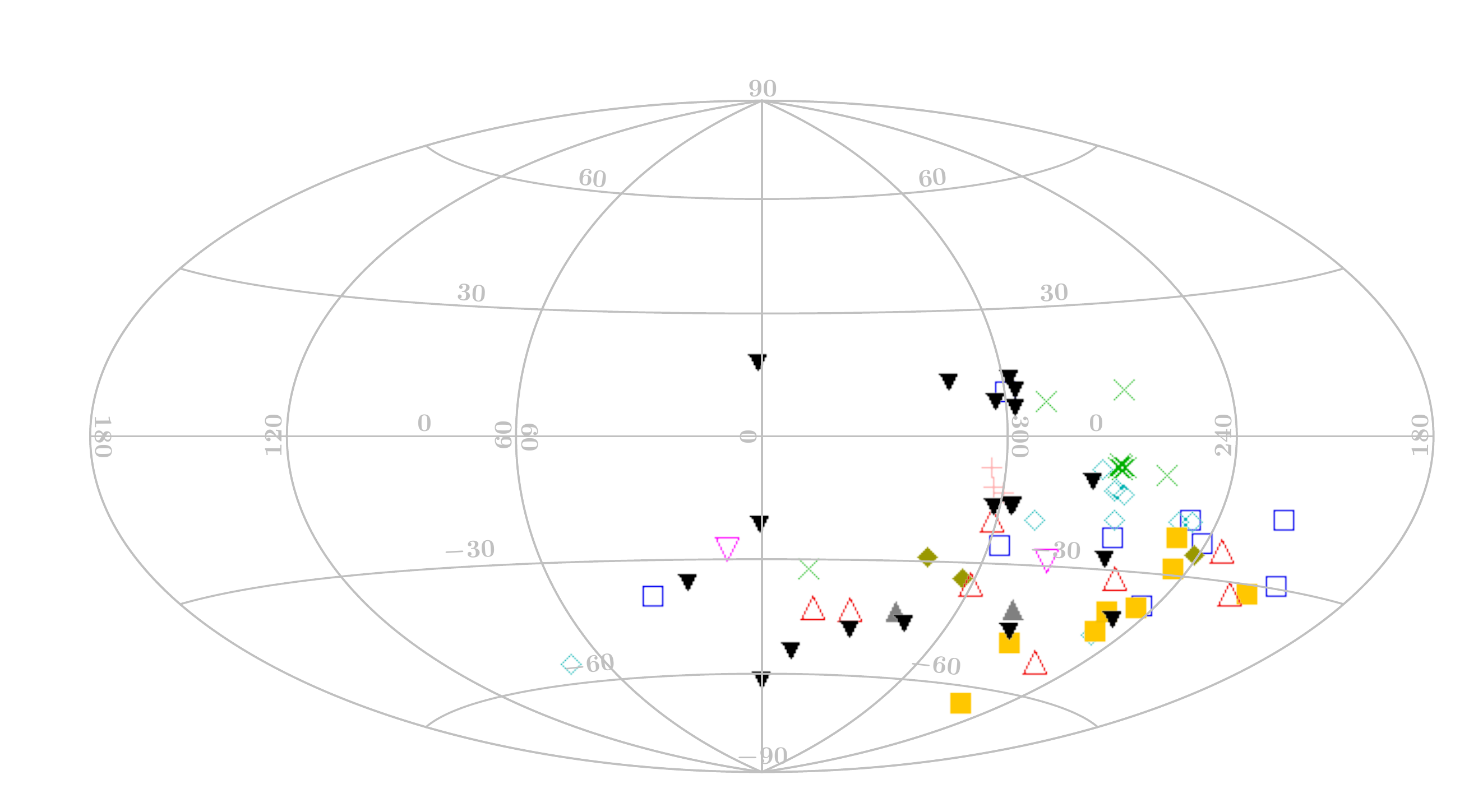}
	\includegraphics[width=0.12\linewidth]{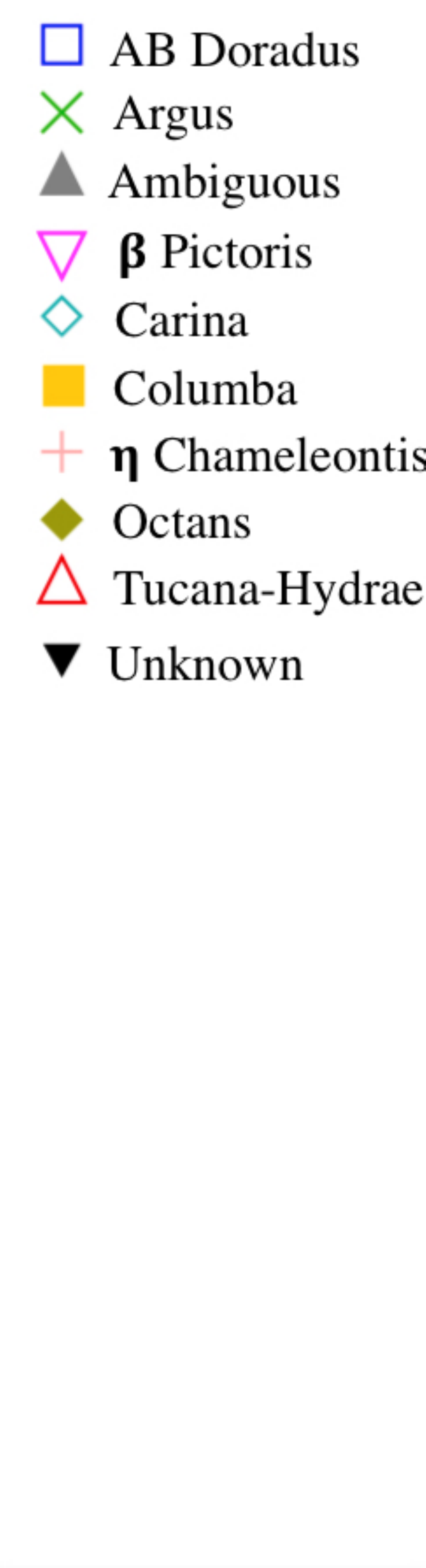}
	\caption{The panel shows the sky map of the detected members. The symbols on the right superior corner shows the detected groups. The map is on galactic coordinates. Unknown refers to stars with no established membership to a YNMG and ambiguous refers to stars that may belong to two or more YNMG. }
	\label{moving-groups}
\end{figure*}

\begin{figure*}[h]
	\centering
	\includegraphics[width=0.7\linewidth]{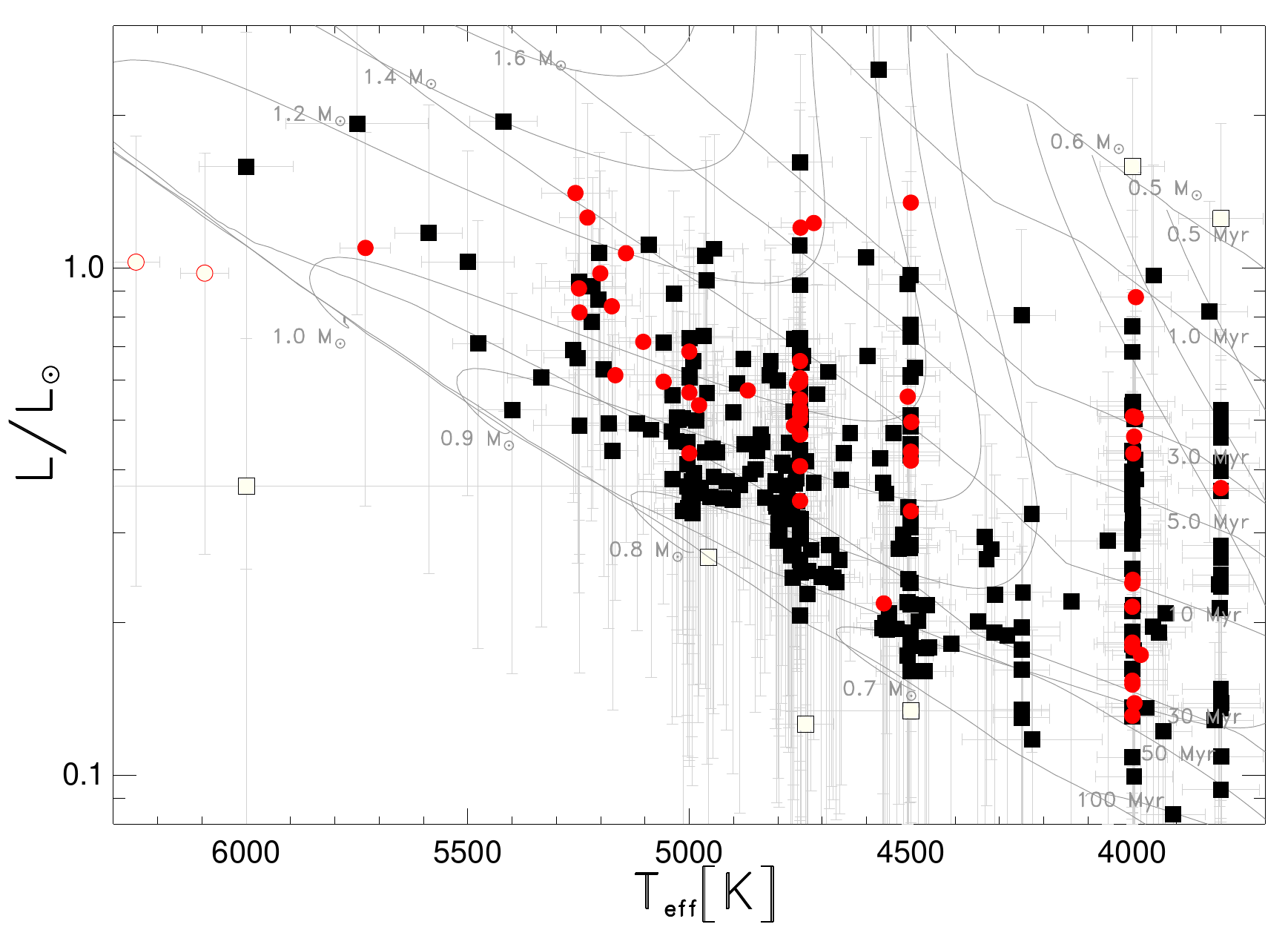}
	\caption{HR diagram of the candidate and member sample. The red points represent members of YNMGs already reported in literature (see Section \ref{s:mg}). The black points represent our sample of possible candidates to belong to YNMGs. The error bars were calculated from the error propagation of the stellar parameters described on section \ref{s:diagrams-HR}. The labeled evolutionary tracks and isochrones are from \cite{Bressan2012} and \cite{Marigo2017}, respectively. The open symbols indicate the sources for which neither ages nor masses were estimated because they lie outside the models grid. }
	\label{HR}
\end{figure*}

\subsection{HR diagrams \label{s:diagrams-HR}}

The remaining  571 stars identified with our method, may still have kinematics in common even though they are not associated to any known YNMG. For instance, a number of these stars could be unidentified members or new groups. However, the list may also contain a fraction of contaminant sources. This mainly comes from the fact that chromospheric activity is not exclusive of young stars; some types of evolved sources in the sub-giant and giant branch \citep{Oz2018}, and spectroscopic binaries \citep{Fekel2002} may present chromospheric emission and be selected as RAVE CAS. This occurs because chromospherical activity can be present before and after the main sequence phase \citep{Frasca2015}. 
As mentioned in Section 1, our main goal is to select members of recently dispersed young clusters, so we need to know which sources are consistent with young ages (10-10$^2$ Myr).

The original goal of this work is the identification of YNMG member candidates through their kinematic signature, but estimating the individual ages of our candidates to isolate probably young stars, helps us to depurate our sample and to reinforce the results.

Combining the $T_{eff}$ and $A_{v}$ from the RAVE DR5 catalog with the distances from \cite{Luri2018} and using the optical $V$ magnitudes from APASS DR9 \citep{Henden2016}, we constructed the HR diagram for the CMS candidate sample. This diagram allowed us to estimate the ages and masses of the candidates working with a method similar to that used by \cite{Suarez2017}, which basically interpolates the $\mathrm{T_{eff}}$ and $\mathrm{L_{bol}}$ into the stellar models from \cite{Bressan2012} and \cite{Marigo2017} to estimate masses and ages, as well as their uncertainties.
For our purposes, in order to remove most of the contamination by evolved sources, we limited our candidate sample to those stars with  $\mathrm{M_V}\geqslant 4.5$ mag. 
After this cutoff, we selected those stars younger than 100 Myr.
The resulting clean sample contains 290 candidates, which represents the $\sim$50\% of the sample of YNMG member candidates. From the remain 280 removed stars, 70 stars are $H\alpha$-emitters. This percentage is consistent with the fraction of CAS in RAVE with high H$\alpha$ emission (see Section \ref{s:sample}). This show that the use of HR diagrams allows to depurate the YNMG member candidate selection.

In Figure \ref{HR} we show the HR diagram for the members and candidates of YNMGs.
We can see that all of member and candidates lie between the 1 and 100 Myr isochrones from \cite{Marigo2017} and between the 0.5 and 1.6 $M_{\odot}$ evolutionary tracks from \cite{Bressan2012}.

\section{Discussion and Summary \label{s:discussion}}

From our analysis, we found that the kinematic identification of YNMGs in the RAVE CAS sample, directly from the velocity ellipsoid projections, was not satisfactory. This is likely because the sample is not robust enough or, perhaps because the number of CAS sources that we can identify as known members of individual YNMGs is small. It may be possible that for a larger sample of stars, that included sources located in a larger volume, such method could be implemented successfully and its velocity ellipsoid could distinguish between groups with different kinematic signature. 

Precisely this problem was the origin of our idea of the alternative method, the CMS we present in this paper. Our method uses directly the orientation of the velocity vectors in the velocity space to identify groups of stars with a common kinematic signature. This made our method distinc from others typically used.

While the idea behind the CMS is relatively simple, we showed that it can be a reliable tool, useful for the identification of kinematically cold groups. Moreover, the CMS appears to work well with small groups in relatively small samples like the RAVE CAS. In this sense, our method is not exclusive for groups of young stars. Our CMS, in principle, can be applied to any catalog of sources that have reliable observational parameters to construct PV6 vectors. This makes our method very well suited for other applications. We applied our method to samples of known groups and detected those overdensities on the $(\theta, \phi)$ map.

With our method, we successfully identified 75 members of YNMGs previously reported in the literature (see appendix \ref{App3}). For the remaining sources in the sample, our analysis of the HR diagram, indicates that a significant number of sources are consistent with ages between 1 and 100 Mys, indicating that the RAVE CAS sample possibly contains dozens of young stars that belong to known or even new YNMG in the solar neighborhood. Follow-up work should focus on determination of membership for the stars in Table \ref{tab:results}, based on youth signatures from spectroscopy.

\acknowledgments
\begin{center}
ACKNOWLEDGMENTS.
\end{center}
VRP and GS acknowledge support from a graduate studies fellowship from CONACYT/UNAM Mexico. VRP, CRZ and GS acknowledges support from program UNAM-DGAPA-PAPIIT IN108117, Mexico. LA acknowledges support from UNAM-DGAPA-PAPIITT IG100125 project.

This paper makes use of RAVE data. Funding for RAVE has been provided by: the Australian Astronomical Observatory; the Leibniz-Institut fuer Astrophysik Potsdam (AIP); the Australian National University; the Australian Research Council; the French National Research Agency; the German Research Foundation (SPP 1177 and SFB 881); the European Research Council (ERC-StG 240271 Galactica); the Istituto Nazionale di Astrofisica at Padova; The Johns Hopkins University; the National Science Foundation of the USA (AST-0908326); the W. M. Keck foundation; the Macquarie University; the Netherlands Research School for Astronomy; the Natural Sciences and Engineering Research Council of Canada; the Slovenian Research Agency; the Swiss National Science Foundation; the Science \& Technology Facilities Council of the UK; Opticon; Strasbourg Observatory; and the Universities of Groningen, Heidelberg and Sydney. The RAVE web site is at https://www.rave-survey.org.

This publication makes use of data products from the Two Micron All Sky Survey, which is a joint project of the University of Massachusetts and the Infrared Processing and Analysis Center/California Institute of Technology, funded by the National Aeronautics and Space Administration and the National Science Foundation.

This work has made use of data from the European Space Agency (ESA) mission
{\it Gaia} (\url{https://www.cosmos.esa.int/gaia}), processed by the {\it Gaia}
Data Processing and Analysis Consortium (DPAC,
\url{https://www.cosmos.esa.int/web/gaia/dpac/consortium}). Funding for the DPAC
has been provided by national institutions, in particular the institutions
participating in the {\it Gaia} Multilateral Agreement.

Some figures in this paper, as well as distance estimates were obtained with the fabulous TOPCAT software, which is described in \cite{topcat}.

\software{Python NumPy, IDL, TopCat}

\bibliographystyle{yahapj}

\appendix

\section{Uncertainty cone} 
\label{App1}
The following error propagation is a linear approximation to obtain the suitable $\alpha$ aperture for the CMS.
Starting from the definitions of $\theta$ and $\phi$ from the transformation of the velocity space
\begin{equation*}
\theta = \arctan{(\frac{W}{\sqrt{U^2 + V^2 }})},
\end{equation*}

\begin{equation*}
\phi = \arctan{(\frac{V}{U})},
\end{equation*}

we calculated the error propagation at first order:
\begin{equation}
\delta \phi = \delta U (\frac{UW}{(\sqrt{U^2 + V^2}) (U^2 + V^2 + W^2)}) + \delta V (\frac{VW}{(\sqrt{U^2 + V^2}) (U^2 + V^2 + W^2)}) + \delta W (\frac{\sqrt{U^2 + V^2}}{U^2 + V^2 + W^2})
\end{equation}
and
\begin{equation}
\delta \theta = \delta U (\frac{U}{\sqrt{U^2 + V^2}}) + \delta V (\frac{U}{\sqrt{U^2 + V^2}})
\end{equation}
Considering that $\delta \alpha$ is the equivalent radius of the segment formed by $\delta \theta \times \delta \phi$, then $\delta \alpha$ is calculated:
\begin{equation}
\delta \alpha = \sqrt{\frac{|\delta \theta| \times |\delta \phi|}{\pi}}
\end{equation}
\\

\section{YNMG detected members in the CAS RAVE Sample} 
\label{App3}

\startlongtable
\begin{center}
\begin{deluxetable*}{lccccccccccccc}
\tabletypesize{\scriptsize}
\tablecolumns{13}
\tablewidth{0pt}
\tablecaption{Main parameters for the members of YNMGs detected with the CMS.}
\tablehead{
\colhead{ID 2MASS}  & 
\colhead{RA}  &
\colhead{DEC}  &
\colhead{Vmag \tablenotemark{a}}  &
\colhead{Jmag \tablenotemark{a}}  &
\colhead{$\mu_{\alpha}$\tablenotemark{b}}  &
\colhead{$\mu_{\delta}$\tablenotemark{b}}  &
\colhead{$V_{R}$ \tablenotemark{c}}  &
\colhead{$T_{eff}$} \tablenotemark{c} &
\colhead{$L_{bol}$\tablenotemark{d}}  &
\colhead{Mass \tablenotemark{d}}   &
\colhead{Age \tablenotemark{d}}   &
\colhead{YNMG\tablenotemark{e}}  &
\\
\colhead{} &
\colhead{} &
\colhead{} &
\multicolumn{2}{c}{[Mag.]} &
\multicolumn{2}{c}{[mas/yr]} &
\colhead{[$km/s$]} &
\colhead{[K]} &
\colhead{} &
\colhead{[$M_{\odot}$]} &
\colhead{[$10^6$ yr]} \\
}
\startdata
  23093711-0225551 & 23:09:37.10 & -02:25:55.0 & 10.933 & 8.567 & 60.842 & -45.963 & -11.21 & 4000.0 & 0.215 & 0.734 & 12.27 & CAR\\
  23215251-6942118 & 23:21:52.50 & -69:42:12.0 & 10.008 & 8.657 & 41.368 & -32.023 & 5.6 & 5248.0 & 0.816 & 1.045 & 19.08 & Unknown\\
  05023042-3959129 & 05:02:30.40 & -39:59:13.0 & 10.654 & 8.727 & 35.104 & -23.752 & 26.329 & 4561.0 & 0.218 & 0.752 & 39.81 & ABDMG\\
  22463348-3928451 & 22:46:33.50 & -39:28:45.0 & 9.528 & 8.21 & 75.105 & -3.24 & -1.64 & 5175.0 & 0.839 & 1.08 & 16.49 & Unknown\\
  05332558-5117131 & 05:33:25.60 & -51:17:13.0 & 11.805 & 8.986 & 42.755 & 26.101 & 18.825 & 4000.0 & 0.153 & 0.706 & 35.37 & THA\\
  03241504-5901125 & 03:24:15.00 & -59:01:13.0 & 12.108 & 9.547 & 43.729 & 7.913 & 19.376 & 4000.0 & 0.239 & 0.734 & 10.04 & COL\\
  05451623-3836491 & 05:45:16.30 & -38:36:49.0 & 11.016 & 9.561 & 10.622 & 7.65 & 30.688 & 5230.0 & 1.256 & 1.245 & 10.85 & COL\\
  08430040-5354076 & 08:43:00.40 & -53:54:08.0 & 11.099 & 9.755 & -23.293 & 22.849 & 12.148 & 5249.0 & 0.91 & 1.092 & 16.57 & ARG\\
  13544209-4820578 & 13:54:42.10 & -48:20:58.0 & 11.024 & 9.285 & -31.868 & -23.39 & 0.338 & 5000.0 & 0.683 & 1.047 & 16.27 & Unknown\\
  11594226-7601260 & 11:59:42.30 & -76:01:26.0 & 11.139 & 9.14 & -41.025 & -6.19 & 14.435 & 3996.0 & 0.465 & 0.735 & 3.491 & ECh\\
\enddata

\tablenotetext{a}{
Photometry of APASS DR9 \citep{Henden2016}}

\tablenotetext{b}{
Taken from GAIA DR2 \citep{GAIADR2}}

\tablenotetext{c}{
Taken from \cite{kunder2017}}

\tablenotetext{d}{
Estimated with a method like \cite{Suarez2017}}

\tablenotetext{e}{
Comparison with catalog provided by M. Rodriguez et al. in prep. Unknown refers to stars with no established membership to a YNMG.}
\label{tab:members}
\end{deluxetable*}
\end{center}

\section{YNMG Candidates in the CAS RAVE Sample} 
\label{App2}

\startlongtable
\begin{center}
\begin{deluxetable*}{lcccccccccccc}
\tabletypesize{\scriptsize}
\tablecolumns{12}
\tablewidth{0pt}
\tablecaption{Main Parameters for the candidates detected with the CMS.}
\tablehead{
\colhead{ID 2MASS}  & 
\colhead{RA}  &
\colhead{DEC}  &
\colhead{Vmag \tablenotemark{a}}  &
\colhead{Jmag \tablenotemark{a}}  &
\colhead{$\mu_{\alpha}$\tablenotemark{b}}  &
\colhead{$\mu_{\delta}$\tablenotemark{b}}  &
\colhead{$V_{R}$ \tablenotemark{c}}  &
\colhead{$T_{eff}$\tablenotemark{c}}  &
\colhead{$L_{bol}$\tablenotemark{d}}  &
\colhead{Mass \tablenotemark{d}}   &
\colhead{Age \tablenotemark{d}}   &
\\
\colhead{} &
\colhead{} &
\colhead{} &
\multicolumn{2}{c}{[Mag.]} &
\multicolumn{2}{c}{[mas/yr]} &
\colhead{[$km/s$]} &
\colhead{[K]} &
\colhead{} &
\colhead{[$M_{\odot}$]} &
\colhead{[$10^6$ yr]} \\
}
\startdata
  23123243-0240516 & 23:12:32.46 & -02:40:51.9 & 12.915 & 10.8 & 17.44 & -4.568 & 11.995 & 4000.0 & 0.21 & 0.732 & 13.05\\
  21323568-5558015 & 21:32:35.70 & -55:58:01.6 & 12.473 & 10.987 & -18.23 & -12.874 & 69.72 & 5750.0 & 1.924 & 1.238 & 14.73\\
  23581157-3850073 & 23:58:11.58 & -38:50:07.5 & 11.689 & 10.104 & -28.779 & -30.763 & 4.097 & 4964.0 & 0.432 & 0.879 & 29.25\\
  00220533-4050257 & 00:22:05.34 & -40:50:25.7 & 13.19 & 11.444 & -5.68 & -19.514 & 4.565 & 5000.0 & 0.727 & 1.072 & 15.12\\
  02523097-5447531 & 02:52:30.99 & -54:47:53.3 & 13.295 & 11.19 & 0.693 & -30.811 & -0.92 & 4500.0 & 0.191 & 0.717 & 52.92\\
  00263489-6545359 & 00:26:34.90 & -65:45:36.0 & 11.289 & 9.611 & 27.489 & 4.21 & -14.657 & 4755.0 & 0.321 & 0.829 & 32.38\\
  05213171-3641084 & 05:21:31.73 & -36:41:08.5 & 12.919 & 10.749 & -1.924 & 17.613 & -11.656 & 4000.0 & 0.341 & 0.731 & 5.331\\
  10573417+0048243 & 10:57:34.16 & +00:48:24.2 & 13.533 & 11.348 & -40.667 & 16.654 & -28.338 & 4465.0 & 0.178 & 0.706 & 60.11\\
  12530218-1549546 & 12:53:02.15 & -15:49:54.6 & 14.267 & 11.37 & -3.745 & 17.96 & -5.038 & 3814.0 & 0.128 & 0.704 & 29.81\\
  21480570-0127397 & 21:48:05.71 & -01:27:39.9 & 11.642 & 10.017 & 14.8 & 12.746 & -1.185 & 5003.0 & 0.336 & 0.8 & 43.78\\
 \enddata
\tablenotetext{a}{
Photometry of APASS DR9 \citep{Henden2016}}

\tablenotetext{b}{
Taken from GAIA DR2 \citep{GAIADR2}}

\tablenotetext{c}{
Taken from \cite{kunder2017}}

\tablenotetext{d}{
Estimated with a method like \cite{Suarez2017}}

\label{tab:results}
\end{deluxetable*}
\end{center}
\section{}

\end{document}